\documentclass[11pt]{article}
\usepackage{amsmath}
\usepackage{amsfonts,amssymb,latexsym}

\begin{document}

\newcommand{\nl}{\nonumber\\}
\newcommand{\nnl}{\nl[6mm]}
\newcommand{\nle}{\nl[-2.5mm]\\[-2.5mm]}
\newcommand{\nlb}[1]{\nl[-2.0mm]\label{#1}\\[-2.0mm]}

\renewcommand{\leq}{\leqslant}
\renewcommand{\geq}{\geqslant}

\newcommand{\bra}[1]{\big{\langle}#1\big{|}}
\newcommand{\ket}[1]{\big{|}#1\big{\rangle}}

\newcommand{\be}{\bes}
\newcommand{\ee}{\ees}
\newcommand{\bes}{\begin{eqnarray}}
\newcommand{\ees}{\end{eqnarray}}
\newcommand{\eens}{\nonumber\end{eqnarray}}

\renewcommand{\/}{\over}
\renewcommand{\d}{\partial}
\newcommand{\no}[1]{{\,:\kern-0.7mm #1\kern-1.2mm:\,}}

\newcommand{\dx}{{d^{d+1}\kern-0.2mm x}}
\newcommand{\dxx}{{d^d\kern-0.2mm x}}
\newcommand{\dk}{{d^{d+1}\kern-0.2mm k}}

\newcommand{\vect}{{\mathfrak{vect}}}
\newcommand{\map}{{\mathfrak{map}}}
\newcommand{\oj}{{\mathfrak{g}}}

\newcommand{\e}{{\mathrm e}}
\newcommand{\ad}{{\mathrm{ad}}}
\newcommand{\tr}{{\mathrm{tr}\,}}

\newcommand{\eps}{\epsilon}
\newcommand{\ww}{\omega}
\newcommand{\la}{\lambda}

\newcommand{\EE}{{\mathcal E}}

\newcommand{\mn}{{\mu\nu}}
\newcommand{\nm}{{\nu\mu}}

\newcommand{\hati}{{\hat\imath}}
\newcommand{\hatj}{{\hat\jmath}}

\newcommand{\cm}{{,\mm}}
\newcommand{\cn}{{,\nn}}
\newcommand{\cmmu}{{,\mm+\hat\mu}}
\newcommand{\cmj}{{,\mm+\hatj}}
\newcommand{\cmnoll}{{,\mm+\hat0}}

\newcommand{\xx}{{\mathbf x}}
\newcommand{\yy}{{\mathbf y}}
\newcommand{\kk}{{\mathbf k}}
\newcommand{\mm}{{\mathbf m}}
\newcommand{\nn}{{\mathbf n}}
\newcommand{\qq}{{\mathbf q}}
\newcommand{\zero}{{\mathbf 0}}

\newcommand{\Tot}{\mathrm{Tot}}
\newcommand{\diag}{\mathrm{diag}}
\newcommand{\undefined}{\mathrm{undefined}}

\newcommand{\J}{{\mathcal J}}
\newcommand{\N}{{\mathcal N}}

\newcommand{\RR}{{\mathbb R}}
\newcommand{\TT}{{\mathbb T}}
\newcommand{\ZZ}{{\mathbb Z}}
\newcommand{\NN}{{\mathbb N}}

\title{{QJT as a Regularization: \break 
Origin of the New Gauge Anomalies}}

\author{T. A. Larsson \\
Vanadisv\"agen 29, S-113 23 Stockholm, Sweden\\
email: thomas.larsson@hdd.se}

\maketitle
\begin{abstract}
QJT is considered as a regularization of QFT, where the fields are
replaced by finite $p$-jets. The regularized phase space is
infinite-dimensional, because not all histories are determined by initial
conditions. Gauge symmetries are not fully preserved by the
regularization, and gauge anomalies arise. These anomalies are of a new
type, not present in QFT. They generically diverge when the regulator is
removed, but can be made finite with a particular choice of field content,
provided that spacetime has at most four dimensions. The field content
appears to include unphysical fields that violate the spin-statistics
theorem.
\end{abstract}

\bigskip

\section{Introduction}

Quantum Jet Theory (QJT) is an approach to quantization of field 
theories, in which not only the fields but also the observer's
trajectory acquires quantum dynamics \cite{Lar04,Lar06,Lar08a}.
The main idea is to replace every quantum field by a jet, which is 
essentially the same thing as a Taylor expansion around some point $q$. 
A Taylor series is observer dependent in the sense that it depends on
the choice of expansion point, which can be identified with the 
observer's position.

There are good reasons to expect that QJT becomes important in the
quantization of gravity. Namely, QJT is a deformation of QFT, where the
deformation parameters are the observer's mass $M$ and charge $e$.
QFT is recovered in the limit $M\to\infty$ (so the observer's position
and velocity commute) and $e\to0$ (so the observer does not disturb the
fields). This limit is readily taken for all interactions except gravity,
where $e = M$. The physical problem with a QFT of gravity is thus that
we tacitly assume both that the observer's inert mass is infinite and 
that his heavy mass is zero.

Historically, QJT grew out of the projective representation theory of
gauge and diffeomorphism algebras, i.e. the multi-dimensional analogues of
affine and Virasoro algebras \cite{Lar98,RM94}. Jets become essential
because it is impossible to construct lowest-energy representations of
these algebras starting from the fields themselves; one must pass to
trajectories in the space of $p$-jets before quantization. The fact that
QJT leads to new anomalies proves that it is substantially different from
QFT, which is positive since QFT is incompatible with gravity.

With this history in mind, it is not surprising that the most striking
feature of QJT is the appearence of new gauge and diff anomalies, which
can not be formulated within QFT. These anomalies were discussed in
\cite{Lar08b}, but that paper was unfortunately unusually opaque, even
for this author. The purpose of the present paper is to	clarify the 
origin of these new anomalies.

In quantum theory, a symmetry may become anomalous if there is no
regularization which respects the symmetry, and the new QJT anomalies
are no exception. QJT suggests a natural
regularization procedure; replace $\infty$-jets by $p$-jets, $p$ finite,
i.e. truncate the Taylor expansions at order $p$. The algebra of gauge
transformations acts in a well-defined manner on $p$-jets, because it can
only lower the order of Taylor coefficients. However, the $p$-jet
regularization is still breaks the symmetry, because the equations of
motions can only be implemented in the ``body'' and not in the ``skin''.
Although a $p$-jet itself only has finitely many components, the $p$-jet
phase space becomes infinite-dimensional, because some histories are not
determined in terms of initial conditions. The new anomalies arise in this
infinite-dimensional ``skin'' of the $p$-jet phase space.

According to popular belief, a gauge anomaly is automatically a sign of
inconsistency. However, this is not correct. E.g., the no-ghost theorem in
string theory asserts that the free string can be quantized without ghosts
in $d \leq 26$ dimensions \cite{GSW87}. In other words, also the
subcritical free string defines a consistent quantum theory, despite its
conformal anomaly. In general, a gauge anomaly turns a classical gauge
symmetry into a quantum global symmetry, which acts on the Hilbert space
instead of reducing it. This may or may not be consistent, depending on
whether the unreduced Hilbert space admits a positive-definite metric
preserved by the anomalous gauge symmetry.

Whereas a finite gauge anomaly may (or may not) be consistent, it seems
obvious that an infinite gauge anomaly renders the theory nonsensical. The
anomalies in the $p$-jet regularization depend on $p$, and diverge in the
field theory limit $p \to \infty$ if the number of spatial dimensions $d
\geq 2$. However, it was observed in \cite{Lar04,Lar06} that the divergent
part could be made to cancel, provided that $d \leq 3$. In this sense, QJT
correctly postdicts that spacetime has $3+1$ dimensions. However, the
cancellations lead to further conditions on the field content, which seem
difficult to reconcile with physics. In the present paper we concentrate
on gauge anomalies in theories of Yang-Mills type, and we find below that,
in addition to the gauge field and matter fermions, one must add a field
which violates the spin-statistics theorem: a fermion with a second-order
equation of motion. It is unclear how this result should be interpreted.
However, it still represents progress compared to \cite{Lar06},  where 
not even unphysical solutions to the consistency conditions were known.

\section{ Free scalar field }

\subsection{ Fields }
\label{ssec:fields}

The basic idea behind QJT is best described with a simple example: a free 
scalar field $\phi(x) = \phi(\xx, t)$ in $(d+1)$-dimensional spacetime.
As usual, spacetime indices are denoted by Greek letters $\mu$, $\nu$, 
and spatial indices by Latin letters $i$, $j$: $x = (x^\mu) \in \RR^{d+1}$ 
has the spacetime decomposition $x = (\xx, x^0)$, where $\xx = (x^i)$; 
in this section, $t = x^0$ denotes the time coordinate.

The action reads
\be
S = {1\/2} \int \dx\ \Big( (\d_0\phi)^2 - (\nabla\phi)^2 - \ww^2 \phi^2
\Big).
\ee
The mass is denoted by $\ww$ to avoid confusion with multi-indices 
introduced below. The equations of motion are of the form $\EE(\xx,t) = 0$,
where
\be
\EE = -{\delta S\/\delta\phi} = \d_0^2\phi - \nabla^2\phi + \ww^2\phi.
\label{Efield}
\ee
The phase space may be identified with the space of solutions of the
equations of motion \cite{HT92}. Since the equations (\ref{Efield}) are 
second order, the solutions depends on $\phi(\xx,0)$ and 
$\d_0\phi(\xx,0) = \pi(\xx)$. Alternatively, we can coordinatize phase
space by the Fourier modes $a_\kk$ and $a^\dagger_\kk$, because a general
solution is of the form
\be
\phi(\xx,t) = \int \dk\,\Big( a_\kk \e^{i(k_0t+\kk\cdot\xx)}
+ a^\dagger_\kk \e^{i(-k_0t+\kk\cdot\xx)} \Big),
\label{phiaa}
\ee
where $k_0 = \ww_\kk \equiv \sqrt{k^2 + \ww^2}$.
The relation between the two phase space bases is obviously
\bes
\phi(\xx,0) &=& \int \dk\,(a_\kk + a^\dagger_\kk)
 \e^{i\kk\cdot\xx}, 
\nlb{PS}
\d_0\phi(\xx,0) &=& i \int \dk\,k_0 (a_\kk - a^\dagger_\kk) 
 \e^{i\kk\cdot\xx}.
\eens

\subsection{ Infinite jets }

Next we reformulate dynamics of the free scalar field in jet space. By 
definition,
a $p$-jet is an equivalence class of functions; two functions belong to
the same class if their derivatives up to order $p$, evaluated at some
given point $\qq$, are the same. Locally, a $p$-jet has a unique
representative which is a polynomial of order at most $p$, namely the
truncated Taylor series around the point $q$. We may and will therefore
identify $p$-jets with Taylor expansions truncated at order $p$; an
infinite jet is hence a Taylor series where the sum continues to infinite
order.

The field $\phi(\xx,t)$ corresponds to the $\infty$-jet with components
\break $\{\phi_\cm(t), q^i(t)\}$ via the Taylor expansion
\be
\phi(\xx,t) = \sum_\mm {1\/\mm!}\,\phi_\cm(t)\,(\xx-\qq(t))^\mm.
\label{Taylor}
\ee
We employ standard multi-index notation: 
\be
\mm = (m_1, m_2, ..., m_d), \qquad \hbox{all $m_i \geq 0$},
\label{mm}
\ee
is a multi-index with length 
$|\mm| = \sum_{j=1}^d m_j$. The factorial is defined by
$\mm! = m_1! m_2! ... m_d!$ and the power by
$(\xx-\qq)^\mm = (x^1 - q^1)^{m_1} (x^2 - q^2)^{m_2}... (x^d - q^d)^{m_d}$.
Denote by $\mm+\hatj$ the multi-index with one extra in the $j$:th 
position, i.e. 
\be
\mm+\hatj = (m_1, m_2,..., m_j + 1, ..., m_d).
\ee
The Taylor coefficients of the partial derivative field $\d_j\phi$ are 
$\phi_\cmj$.

We assume that the time coordinate is given by
\be
x^0 = q^0(t) = t.
\label{xqt}
\ee
The time derivative acts as
\be
\d_0\phi(\xx,t) = \d_t \phi
= \sum_\mm {1\/\mm!} (\,\dot\phi_\cm(t) 
- \sum_{j=1}^d \dot q^j\phi_\cmj\,) \,(\xx-\qq(t))^\mm.
\ee
For convenience, we define coefficients $\phi_\cmnoll$ by
\be
\d_0\phi(\xx,t) \equiv 
\sum_\mm {1\/\mm!}\,\phi_\cmnoll(t)\,(\xx-\qq(t))^\mm.
\ee
Comparison with the previous equation immediately yields
\be
\phi_\cmnoll = \dot\phi_\cm - \sum_{j=1}^d \dot q^j \phi_\cmj.
\ee
The equations of motion take the form $\EE_\cm = 0$, where
\bes
\EE_\cm &=& \sum_{j=0}^d \eta^\mn \phi_{,\mm+\hat\mu+\hat\nu} 
 + \ww^2\phi_\cm 
\nlb{Ejet}
&\equiv& \phi_{,\mm+2\hat0} - \sum_{j=1}^d \phi_{,\mm+2\hatj} 
 + \ww^2\phi_\cm,
\eens
where $\eta^\mn = \diag(1, -1, ..., -1)$ is the Minkowski metric and
\bes
\phi_{,\mm+2\hat0} &=& \ddot\phi_\cm - 2\sum_{j=1}^d \dot q^j\dot\phi_\cmj 
\nlb{fm20}
&&- \sum_{j=1}^d \ddot q^j\phi_\cmj 
+ \sum_{i,j=1}^d \dot q^i\dot q^j\phi_{,\mm+\hati+\hatj}.
\eens
As the notation suggests, $\EE_\cm$ are the Taylor coefficients of the
equation of motion field $\EE(\xx,t)$ in (\ref{Efield}). 

The equation $\EE_\cm = 0$ can be written as $\ddot\phi_\cm = ...$, and is
thus a second-order equation for the Taylor coefficients $\phi_\cm(t)$. A
solution is fully specified by the Cauchy data $\phi_\cm(0)$ and
$\dot\phi_\cm(0)$. Since phase space can be identified with the space of
solutions, a basis for phase space is given by $\phi_\cm(0)$ and
$\pi_\cm(0) = \phi_\cmnoll(0)$. The phase space is infinite-dimensional,
because the multi-index $\mm \in \NN^d$ can take infinitely many values.
This is expected, because an infinite jet contains essentially the same
information as the field itself, modulo assumptions about convergence of
the Taylor series (\ref{Taylor}). To find the relation to the Fourier
basis in (\ref{phiaa}), we expand the exponential as
\be
\e^{i\kk\cdot\xx} = \e^{i\kk\cdot\qq} 
\sum_\mm {1\/\mm!}\,(i\kk)^\mm\,(\xx - \qq)^\mm.
\ee 
We find
\bes
\phi_\cm(0) &=& \int \dk\,(a_\kk + a^\dagger_\kk) \e^{i\kk\cdot\qq}(i\kk)^\mm, 
\nle
\pi_\cm(0) &=& \phi_\cmnoll(0) =
 \int \dk\,ik_0 (a_\kk - a^\dagger_\kk) \e^{i\kk\cdot\qq}(i\kk)^\mm.
\eens
To make sense of the divergent integrals over $\kk$ involves subtleties
which are ignored.

\subsection{ Finite $p$-jets }

QJT has a natural built-in regularization method: truncate from infinite
jets to $p$-jets, $p$ finite. This means that the Taylor series 
(\ref{Taylor}) is truncated at order $p$, i.e.
\be
\phi(\xx,t) \approx \sum_{|\mm|\leq p} 
 {1\/\mm!}\,\phi_\cm(t)\,(\xx-\qq(t))^\mm.
\ee

A basis for the space of all histories in $p$-jet space consists of
$q^i(t)$ and $\phi_\cm(t)$ for all $\mm$ such that $|\mm|\leq p$. This
suggests that the phase space, i.e. the space of histories which solve the
equations of motion, should be spanned by $\phi_\cm(0)$ and 
$\pi_\cm(0) = \phi_\cmnoll(0)$ with $|\mm|\leq p$. Hence the regulated
phase space should be finite-dimensional, and in fact the dimension should
equal $2{d+p \choose d}$.

However, except for case $d=0$, i.e. ordinary quantum mechanics, this 
expectation is wrong. When $d = 0$, we have the phase space of the 
harmonic oscillator, which indeed is two-dimensional and spanned e.g. by 
the vectors $\phi(0)$ and $\pi(0)$. To see what goes wrong in higher
dimensions, we return to the Taylor expansion of the equations of 
motion. The second term in (\ref{Ejet}) involves the term
$\phi_{,\mm+2\hatj}$, which is only defined for $|\mm| \leq p-2$ since
$\phi_\cm$ is only defined for $|\mm| \leq p$. Therefore, the equations 
of motion for $\phi_\cm$ with $|\mm| = p-1$ or $p$ are undefined. The 
situation is the same for the time derivative $\phi_{,\mm+2\hat0}$, which
according to (\ref{fm20}) also depends on spatial derivatives of order
$|\mm|+2$.

The correct equations of motion read
\be
\EE_\cm = \begin{cases}
\phi_{,\mm+2\hat0} - \sum_{j=1}^d \phi_{,\mm+2\hatj} + \ww^2\phi_\cm, 
& |\mm| \leq p-2 \\
\undefined & |\mm| = p-1, p
\end{cases}
\label{Epjet}
\ee
One may imagine introducing some dynamics for the top modes, e.g.
$\EE_\cm \equiv 0$ for $|\mm| = p-1, p$. However, such an assumption would
be incorrect, as it would invalidate the solutions (\ref{phiaa}). The
correct treatment is to leave the dynamics undefined for the top modes.
This means that the equations of motion do not fully specify the histories
$\phi_\cm(t)$ in terms of data living at $t=0$. 

The full $p$-jet phase space, i.e. the space of $p$-jet histories which
solve the equations of motion (\ref{Epjet}), is spanned by the basis
\bes
\phi_\cm(0), \pi_\cm(0) = \phi_\cmnoll(0), &\qquad& |\mm| \leq p-2, \nle
\phi_\cm(t), \quad \forall t \in \RR, &\qquad& |\mm| = p-1, p.
\eens
The $p$-jet phase space is infinite-dimensional because the equations
of motion are unable to determine some histories in terms of data living
at $t=0$. 

To facilitate further discussion, we define\footnote{ The concepts
are called body and skin rather than the perhaps more natural terms bulk 
and shell, because the latter have other meanings as well, whereas the 
former appear to be unused.}
\begin{itemize}
\item
The {\bf body} of a $p$-jet consists of the components $\phi_\cm$ such that
the corresponding component $\EE_\cm$ is defined.
\item
The {\bf skin} of a $p$-jet is the complement of the body, i.e. the
components $\phi_\cm$ such that $\EE_\cm$ is undefined.
\item
The {\bf thickness} of the skin is $n$, if the body consists of $\phi_\cm$
with $|\mm| \leq p-n$. For theories without gauge symmetries, the 
thickness is equal to the order of the equations of motion.
\item
The terms body and skin are used about the $p$-jet phase space as well,
to denote the subspace spanned by the body and skin of the $p$-jet.
\end{itemize}
In particular for the scalar field, a $p$-jet component $\phi_\cm$ belongs
to the body if $|\mm| \leq p-2$, it belongs to the skin if $|\mm| = p-1$
or $p$, and the thickness of the skin equals two.

We can now rephrase the main observation of this subsection: the body
of the $p$-jet phase space is finite-dimensional, as expected by
truncation from the $\infty$-jet phase space, but the skin is 
infinite-dimensional, because the equations of motion do not fix the skin
of a $p$-jet in terms of initial data. The infinite-dimensional skin
turns out to be the source of anomalies.

\subsection{ BV-BRST }
\label{ssec:BV1}

The BV (Batalin-Vilkovisky) method \cite{HT92} gives a cohomological
construction of the phase space from arbitrary histories. It underlies
the Manifestly Covariant Canonical Quantization (MCCQ) programme in
\cite{Lar04,Lar06}. Although not without problems, it clarifies some 
aspects of the $p$-jet phase space; in particular, it clearly exhibits
the separation between body and skin.

Return to the field formulation in subsection \ref{ssec:fields}.
In addition to the bosonic fields $\phi(\xx,t)$, we introduce fermionic
antifields $\phi^*(\xx,t)$. Define the BV-BRST operator $\delta$ by
\bes
\delta \phi &=& 0, \nle
\delta \phi^* &=& \EE \ =\ \Box\phi + \ww^2 \phi.
\eens
The cohomology groups are
\bes
H^0(\delta) &=& C(\phi) / \N \nle
H^i(\delta) &=& 0, \qquad \hbox{if $k > 0$},
\eens
where $C(\phi)$ is the space of smooth functionals over $\phi$, and $\N$
is the ideal generated by $\EE$. The BV complex thus yields a 
resolution of the space of functionals over phase space, which consists
of histories that solve the equations of motion $\EE=0$ \cite{HT92}.

In MCCQ, we introduce the canonical momenta in the history phase
space, satisfying the canonical commutation relations (CCR)
\bes
[\phi(\xx,t), \pi(\xx',t')] &=& i\delta(\xx-\xx') \delta(t-t'), 
\nlb{fieldCCR}
\{ \phi^*(\xx,t), \pi^*(\xx',t')\} &=& \delta(\xx-\xx') \delta(t-t').
\eens
Note that the CCR in history space are instantaneous; the RHS is 
proportional to $\delta(t-t')$. We can now rewrite the BV-BRST operator
as a bracket; for any functional $F(\phi,\phi^*)$,
$\delta F = [Q,F]$, where
\be
Q = \int \dxx dt\ \EE(\xx,t) \pi^*(\xx,t).
\label{BVQ}
\ee
The main problem with MCCQ is overcounting; the canonical momentum
$\pi$ is not related to $\d_0\phi$. We can overcome this problem by
making the identification $\pi = \d_0\phi$, which amounts to adding
further terms to $\delta$. However, these terms necessarily break the
manifest covariance which was a main motivation.

In the space of $p$-jet histories, the BV-BRST operators becomes
\bes
\delta \phi_\cm = 0, &\qquad& |\mm| \leq p, \nle
\delta \phi^*_\cm = \EE_\cm, &&|\mm| \leq p-2.
\eens
Since $\EE_\cm$ is only defined for the body of the $p$-jet,
so is the antifield $\phi^*_\cm$. Introduce canonical momenta that
obey the instantaneous CCR
\bes
[\phi_\cm(t), \pi^\cn(t')] &=& 
 i \delta^\nn_\mm \delta(t-t'), 
\nlb{phiCCR}
\{\phi^*_\cm(t), \pi_*^\cn(t')\} &=& 
 \delta_\mm^\nn \delta(t-t').
\eens
The BV-BRST operator can now be written as a bracket $\delta F = [Q,F]$,
where
\be
Q = \sum_{|\mm|\leq p-2} \int dt\ \EE_\cm(t)\pi_*^\cm(t).
\label{Qjet}
\ee
There are two things to note:
\begin{itemize}
\item
The sum only runs over $\mm$ in the body, because the antifield 
$\phi^*_\cm$ and its canonical momentum $\pi_*^\cn$ are only defined there.
\item
$Q$ is already normal ordered, because $\EE_\cm$ is independent of the
antifield. This will no longer be true in the  presence of gauge
symmetries.
\end{itemize}

A jet does not only consist of the Taylor coefficients $\phi_\cm$, but also
of the expansion point $\qq$, which can be thought of as the observer's
position. To fully specify jet dynamics, we must thus introduce some
equations of motion for this quantity as well. For simplicity, we 
equip $q^i(t)$ with the dynamics of a free relativistic point particle
with mass $M$, described by the action
\be
S = -M \int dt\ \sqrt{1 - \dot\qq^2}.
\label{Sq}
\ee
This leads to the equations of motion
\be
\EE^i_q = {d\/dt}\Big( { \dot q^i\/ \sqrt{1 - \dot\qq^2}} \Big) = 0.
\ee
To implement this in cohomology, we introduce antifields $q^i_*(t)$ and
posit that the BV-BRST operator acts as
\bes
\delta q^i &=& 0, \nle
\delta q^i_* &=& \EE^i_q.
\eens

\section{ Reparametrizations }

\subsection{ Fields }

The Taylor expansion (\ref{Taylor}) is obtained from the spacetime jet
\bes
\phi(x) &=& \sum_m {1\/m!}\, \phi_{,m}\, (x - q)^m \nle
&=& \sum_{m_0}\sum_\mm {1\/m_0!\mm!}\, \phi_{,(m_0, \mm)}
 \,(x^0 - q^0)^{m_0} (\xx - \qq)^\mm,
\eens
by setting $x^0 = q^0 = t$, cf (\ref{xqt}). In this section we relax this 
condition, and only assume that there is some function $q^0(t)$ such that
$x^0 = q^0(t)$,
whereas $q^0(t) \neq t$ in general. This function is not completely 
arbitrary, but is assumed to be everywhere smooth and invertible.
Since the condition $x^0 = q^0$ still holds, the $m_0$ dependence in
the Taylor series disappears.

We now have two types of time-like coordinates: the physical time $x^0$,
which appears directly in the equations of motion, and the time parameter
$t$, which is not observable. They are related by 
\bes
x^0 = q^0(t), \qquad t = \tau(x^0).
\label{par}
\ees
Consequently, there are two types of time derivatives:
\be
\d_t \phi = \dot q^0 \d_0 \phi, \qquad
\d_0 \phi = \d_0 \tau \d_t \phi.
\ee
Because the functions $q^0(t)$ and $\tau(x^0)$ are each other's inverses,
$\dot q^0 \d_0 \tau = 1$.

The time parameter $t$ is a non-observable gauge variable. The theory
is invariant under infinitesimal reparametrizations $t \mapsto t + f(t)$.
The field transforms as
\bes
\phi(\xx,t) &\mapsto& \phi(\xx, t-f(t)) \nl
&=& \phi(\xx,t) - f(t) \d_t\phi(\xx,t) \\
&\equiv& \phi(\xx,t) + L_f\phi(\xx,t).
\eens
For simplicity we will only consider reparameterizations which leave the
surface $t = 0$ invariant. Hence we demand that
\be
f(0) = 0.
\ee
This is not an essential restriction, but makes preservation of the
phase space spanned by (\ref{PS}) manifest.
The gauge generators $L_f$ satisfy the Witt algebra
\be
[L_f, L_g] = L_{[f,g]}, \qquad
[f, g] = f \dot g - g \dot f.
\label{Witt}
\ee
Reparametrizations act as
\bes
[L_f, \phi] &=& - f\d_t\phi, \nl
{[}L_f, \d_\mu\phi] &=& - f\d_t\d_\mu\phi, 
\label{LfField}\\
{[}L_f, \EE] &=& - f\d_t\EE.
\eens
In other words, the field $\phi$ and its derivatives w.r.t. physical
coordinates transform as fields with weight $0$ under reparametrizations,
which is necessary because otherwise the equations of motion would not
transform homogeneously. The observer's trajectory and its time derivative
transform as
\bes
{[}L_f, q^\mu] &=& - f\dot q^\mu, 
\nlb{Lfq}
{[}L_f, \dot q^\mu] &=& - f\ddot q^\mu - \dot f \dot q^\mu 
= -{d\/dt}(f \dot q^\mu).
\eens
Hence $q^\mu$ also has weight $0$, whereas
$\dot q^\mu$ is a density with weight $+1$. 
The reparametrization algebra admits the off-shell realization
\be
L_f = i\iint dt\dxx\ f(t)\d_t\phi(\xx,t)\pi(\xx,t),
\ee
where $\pi(\xx,t) = -i\delta/\delta\phi(\xx,t)$ satisfies
(\ref{fieldCCR}). The representation defined by (\ref{LfField}) acts in 
a non-trivial way on general histories, but preserves the phase space 
spanned by $\phi(\xx,0)$ and $\d_0\phi(\xx,0)$, because we assumed that 
$f(0) = 0$. Reparametrizations generate a gauge symmetry under which 
the physical phase space is invariant.

\subsection{ $p$-jets }

It follows from (\ref{LfField}) that the reparametrization algebra acts
in $p$-jet space as
\bes
[L_f, \phi_\cm] &=& -f \dot\phi_\cm, \nl
{[}L_f, \phi_\cmmu] &=& -f \dot\phi_\cmmu, \nle
{[}L_f, \dot\phi_\cm] &=& -f\ddot\phi_\cm - \dot f\dot\phi_\cm 
= -{d\/dt}(f \dot\phi_\cm), \nl
{[}L_f, \EE_\cm] &=& -f\dot\EE_\cm,
\eens
together with the relations written down in (\ref{Lfq}).
Hence it admits the off-shell realization
\be
L_f = i\sum_{|\mm|\leq p} \int dt\ f(t) \dot\phi_\cm(t)\pi^\cm(t) 
+ i\int dt\ f(t)\dot q^\mu(t) p_\mu(t),
\label{LfJet}
\ee
where $\pi^\cm(t) = -i\delta/\delta\phi_\cm(t)$ and 
$p_\mu(t) = -i\delta/\delta q^\mu(t)$; they satisfy the instantaenuous
CCR (\ref{phiCCR}) and 
\bes
[q^\mu(t), p_\nu(t')] &=& i\delta^\mu_\nu \delta(t-t').
\label{qCCR}
\ees

We emphasize that (\ref{LfJet}) is an off-shell realization which is 
valid before the equations of motion have been taken into account. 
Reparametrizations act trivially on the body of the $p$-jet phase
space, for the same reason that they act trivially on the field phase
space: $f(0) = 0$. However, they do not act trivially on the skin, 
which depends on histories $\phi_\cm(t)$ for all $t$. In other words, the
$p$-jet regularization does not preserve the reparametrization gauge
invariance of the original field formulation.

The theory is quantized by introducing a Fock vacuum, which is 
annihilated by negative frequency modes. The body of the phase space
must be polarized in some way, but exactly how this is done is not
important, because this part of the phase space is finite-dimensional.
In contrast, the skin of the phase space is infinite-dimensional and the
choice of polarization is essential. The correct choice is to demand that
negative frequency modes annihilate the vacuum $\ket0$. For simplicity,
assume that the time parameter $t \in S^1$ takes values on the circle.
This assumption is of course unphysical, because it leads to the 
introduction of closed time-like curves, but has some advantages. E.g.,
that we may expand any jet history in Fourier modes:
\be
\phi_\cm(t) = \sum_{\beta = -\infty}^\infty \phi_\cm(\beta) \e^{i\beta t},
\ee
and analogously for the momenta $\phi_\cmnoll(t)$.
We now posit that the vacuum is annihilated by negative-frequency modes: 
\be
\phi_\cm(-\beta)\ket 0 = \phi_\cmnoll(-\beta)\ket 0 = 0, 
\qquad \hbox{ for all $-\beta < 0$}.
\ee
There is some choice how to treat the zero modes with $\beta = 0$, but
exactly how this is done is not essential since they only span a 
finite-dimensional subspace.

The reparametrization generators must be normal ordered to avoid infinite 
contributions after quantization. Hence we must e.g. replace in 
(\ref{LfJet})
\be
\dot\phi_\cm(t)\pi^\cm(t) \mapsto \no{ \dot\phi_\cm(t)\pi^\cm(t)}
= \dot\phi_\cm(t)\pi^\cm_<(t) + \pi^\cm_>(t)\dot\phi_\cm(t),
\ee
where $\pi^\cm_<$ and $\pi^\cm_>$ only runs over negative and positive
Fourier modes, respectively. Moreover, only the skin of the phase space
contributes, because reparametrizations act trivially on the 
finite-dimensional body.  The normal-ordered generators thus read
\be
L_f &=&  \sum_{p-1\leq|\mm|\leq p}
i\int dt\ f(t)\no{ \dot\phi_\cm(t)\pi^\cm(t)}.
\ee
The contribution from the Taylor coefficients comes only from the skin,
which is the difference between the full $p$-jet and the body. We can
therefore write
\be
L_f &=&  L^{(p)}_f - L^{(p-2)}_f, 
\label{LfSkin}
\ee
where
\be
L^{(r)}_f = \sum_{|\mm|\leq r} 
 i\int dt\ f(t)\no{ \dot\phi_\cm(t)\pi^\cm(t)}.
\ee

There is also a contribution from the observer's trajectory 
$q^\mu(t)$ of the form
\be
L^q_f &=& i\int dt\ f(t)\no{ \dot q^\mu(t) p_\mu(t)}.
\ee
The dynamics for $q^i(t)$, which follows from the action 
(\ref{Sq}), reduces the independent degrees of freedom to the 
finite-dimensional space spanned by $q^i(0)$ and $\dot q^i(0)$,
on which reparameterizations act trivially. We discuss this issue
further in subsection \ref{ssec:BV2}.

\subsection{ Reparametrization anomalies }

After normal ordering, the Witt algebra (\ref{Witt}) acquires an
extension and is replaced by the Virasoro algebra
\be
[L_f, L_g] = L_{[f,g]} + {c\/24\pi i} 
\int dt\ \Big( \ddot f(t)\dot g(t) - \dot f(t) g(t) \Big).
\label{Vir}
\ee
As is well known, the contribution from a single bosonic function of $t$
to the central charge is $c = 2$. The number of different multi-indices 
with $|\mm|\leq r$ in $d$ dimensions is ${d+r \choose d}$. In view of
(\ref{LfSkin}), the total central charge for the skin becomes
\be
c_\Tot = 2{d+p \choose d} - 2{d+p-2 \choose d}.
\label{ccBose}
\ee
There are a number of points to observe with this formula.
\begin{itemize}
\item
The reparametrization gauge symmetry becomes anomalous. Classically, we
could eliminate the time parameter $t$ by setting $t = x^0$. Such a gauge
fixing is not allowed after quantization due to the nonzero central charge.
\item
The anomaly originates from the infinite-dimensional skin. The body of
the $p$-jet phase space is finite-dimensional and can not give rise to
anomalies.
\item
When $d = 0$, the central charge (\ref{ccBose}) vanishes: $c = 0$.
\item
When $d = 1$, the central charge is independent of $p$: $c = 4$. The
skin always consists of two functions $\phi_{,p-1}(t)$ and 	
$\phi_{,p}(t)$.
\item
When $d \geq 2$, the central charge diverges in the limit $p\to\infty$.
\item
There is also an additional contribution from the observer's trajectory 
$q^\mu(t)$, which we deal with in the next subsection. However, this 
contribution does not diverge when $p\to\infty$ and is therefore not our
main concern.
\end{itemize}

The crucial observation is that the central charge diverges when $d\geq2$.
We regard this as a sign of inconsistency, which must be avoided if QJT
is to make sense. The rest of this paper is devoted to finding ways to
cancel the infinite parts of anomalies.

\subsection{ BV-BRST }
\label{ssec:BV2}

The main benefit of MCCQ is that it facilitates the counting necessary
to compute anomalies like the one in the previous subsection. When
relaxing the condition $q^0(t) = t$, we turn $q^0$ into a dynamical
variable, and the action (\ref{Sq}) is replaced by
\be
S_q = -M \int dt\ \sqrt{\dot q^\mu \dot q_\mu(t) }.
\ee
Since we have introduced an extra degree of freedom to describe the same
physics, the equations of motion
\be
\EE^\mu_q = {d\/dt}\Big( {\dot  q^\mu\/ \sqrt{\dot q^2}} \Big) = 0
\ee
are redundant:
\be
{\dot q_\mu\/ \sqrt{\dot q^2}}\, \EE^\mu_q 
= {1\/2} {d\/dt}\Big( {\dot  q_\mu \dot q^\mu\/ \dot q^2} \Big) \equiv 0.
\label{qred}
\ee
This can be cast in a more familiar form by noting that the momentum
$p_\mu = M \dot q_\mu/\sqrt{\dot q^2}$ satisfies $p^2 = M^2$, and hence
$dp^2/dt = 0$.

To implement this in cohomology, we introduce fermionice antifields 
$q^i_*(t)$ for the equations of motion $\EE^\mu_q$, bosonic second-order 
antifields $\zeta(t)$ for the redundancy (\ref{qred}), and a fermionic 
ghost $c(t)$ which identifies states related by reparametrizations. 
This gives us the extended phase space of $p$-jet histories, over
which the BV-BRST complex is defined. The BV-BRST operator acts as
\bes
\delta c &=& - c\, \dot c, \nl
\delta \phi_\cm &=& - c\, \dot \phi_\cm, \nl
\delta \phi^*_\cm &=& \EE_\cm - c\, \dot \phi^*_\cm, \nl
\delta q^\mu &=& - c\, \dot q^\mu, 
\label{deltaPar}\\
\delta q_*^\mu &=& \EE^\mu_q - {d\/dt}(c\,q_*^\mu), \nl
\delta \zeta &=& { \dot q_\mu\/ \sqrt{\dot q^2}}\, q_*^\mu 
 - {d\/dt}( c\,\zeta ).
\eens
The reparametrization algebra acts on the extended phase space. Each
quantity in (\ref{deltaPar}) transforms as a density with weights
$-1$ ($c$), $0$ ($\phi_\cm$), $0$ ($\phi^*_\cm$), $0$ ($q^\mu$), $+1$ ($q_*^\mu$),
and $+1$ ($\zeta$), respectively. These weights are reflected in the
terms proportional to $c$ in (\ref{deltaPar}).

The action of the BV-BRST operator can be written as a bracket in the
same way as in subsection \ref{ssec:BV1}. For each quantity in
(\ref{deltaPar}) ($c$, $\phi_\cm$, etc.), we introduce the corresponding
momentum in history space, which is defined by instantaneous 
CCR like (\ref{phiCCR}) or (\ref{qCCR}). 
The BV-BRST charge $Q$ is then defined in analogy with (\ref{Qjet}); for
each relation in (\ref{deltaPar}), we add a term that is linear in
momenta.

However, there is a crucial difference compared to the situation in
subsection \ref{ssec:BV1}: the BRST charge is not automatically normal
ordered. E.g., to implement the relation $\delta\phi_\cm$ the BRST charge
must contain the term
\be
Q_\phi = \sum_{|\mm|\leq p} \int dt\ c(t)\dot\phi_\cm(t) \pi^\cm(t).
\ee
After normal ordering of these ${d+p\choose d}$ terms, $Q_\phi$ is no
longer nilpotent, and reparametrization may fail to be a gauge symmetry
on the quantum level. However, that $Q_\phi^2 \neq 0$ does not necessarily
imply that $Q^2 = 0$; the contributions from different fields to the
anomaly may cancel.

The situation is summarized in the following table.
\[
\begin{array}{|c|c|c|c|c|}
\hline
\hbox{Field} & \hbox{Weight} & \hbox{Order} & \hbox{Parity} & c \\
\hline
c & -1 & 0 & F & -26 \\
\phi_\cm & 0 & p & B & 2{ d+p \choose d } \\
\phi^*_\cm & 0 & p-2 & F & -2{ d+p-2 \choose d } \\
q^\mu & 0 & 0 & B & 2(d+1) \\
q_*^\mu & 1 & 0 & F & -2(d+1) \\
\zeta & 1 & 0 & B & 2 \\
\hline
\end{array}
\]
The columns contain the following information: the type of field, its 
weight under reparametrizations, the maximal order for which is defined 
(only applies to the Taylor coefficients), the Grassmann parity 
(bosonic or fermionic), and the contribution to the central charge.
Adding all contributions to the central charge, we find
\be
c_\Tot = 2{d+p \choose d} - 2{d+p-2 \choose d} - 24,
\ee
which agrees with (\ref{ccBose}) except that the observer's trajectory
and the ghost is no long ignored.

\subsection{ Scalar densities }

The field $\phi(\xx,t)$ does not have to transform as a field, i.e. as
a density with weight zero, under reparametrizations. Instead, it may
transform as a density of weight $\la$. The transformation law
(\ref{LfField}) is then replaced by
\be
[L_f, \phi] &=& - f\d_t\phi - \la\dot f\phi, 
\label{density}
\ee
from which it follows that
\bes
{[}L_f, \d_\mu\phi] &=& - f\d_t\d_\mu\phi - \la\dot f\d_\mu\phi, \nle
{[}L_f, \EE] &=& - f\d_t\EE - \la\dot f \EE.
\eens
However, the weight can only be nonzero if the equations of motion $\EE$
are homogeneous in $\phi$, because otherwise $\EE$ would not transform
homogeneously. The discussion in this paper has been phrased for the 
free scalar field, but the linear equations of motion have not been used
until this point, and the construction goes through also for interacting
theories. However, we must now restrict ourselves to non-interacting 
theories, because nonzero weight is only possible if the 
equations of motion are homogeneous.

Alas, this is not a serious restriction, because we can introduce 
interactions in the presence of several types of fields. Consider
e.g. a scalar field minimally coupled to an electromagnetic field
$A_\mu$. The action reads
\be
S = {1\/2} \int \dx\ \Big( ((\d_\mu + eA_\mu)\phi)^2 - \ww^2\phi^2 \Big).
\label{SAphi}
\ee
The equations of motion 
\be
\EE = (\d_\mu + eA_\mu)^2\phi + \ww^2\phi
\ee
are homogeneous	in $\phi$ but not in $A_\mu$. Thus we may consistently
assume that $\phi$ transforms as a density with any weight $\la$, 
whereas $A_\mu$ must have weight $\la = 0$.

As is well known, the central charge of a single scalar density is
\be
c = 2(1 - 6\la + 6\la^2).
\label{clabos}
\ee
The formula for the total central charge for the skin (\ref{ccBose}) is
replaced by
\be
c_\Tot = 2(1 - 6\la + 6\la^2) \Big( 
{d+p \choose d} - {d+p-2 \choose d} \Big).
\ee
A nonzero weight thus modifies the value of $c_\Tot$, but when $d \geq 2$
it still diverges when $p \to \infty$.

In the presence of additional fields, such as the gauge potential $A_\mu$,
there are additional contributions to the central charge.

\subsection{ Fermions }

Consider a model with a free fermionic field. The action
\be
S = \int \dx\ \bar\psi(i\gamma^\mu\d_\mu - \ww)\psi
\ee
leads to the equations of motion
\bes
\EE &=& i\gamma^\mu\d_\mu\psi - \ww\psi, 
\nle
\bar\EE &=& i\d_\mu\bar\psi\gamma^\mu + \ww\bar\psi.
\eens
Upon passage to $p$-jet space, these equations become
\bes
\EE_\cm = \sum_{\mu=0}^d i\gamma^\mu \psi_\cmmu - \ww\psi_\cm, \nle
\bar\EE_\cm = \sum_{\mu=0}^d i\bar\psi_\cmmu \gamma^\mu 
+ \ww\bar\psi_\cm.
\eens
The equations of motion have order one, so the skin has
thickness one. Moreover, $\EE_\cm$ is linear in $\psi_\cm$ (and
$\bar\EE_\cm$ is linear in $\bar\psi_\cm$), which means that we may 
consistently assume that $\psi_\cm$ transforms as a density of weight $\la$.
The central charge for each fermionic function $\psi_\cm(t)$ is the same as
(\ref{clabos}), up to a sign:
\be
c = -2(1 - 6\la + 6\la^2).
\label{clafer}
\ee
The total central charge comes from the skin of thickness one, and is
given by
\bes
c_\Tot &=& -2(1 - 6\la + 6\la^2)\Big( 
{d+p \choose d} - {d+p-1 \choose d} \Big) \nle
&=&  -2(1 - 6\la + 6\la^2){d+p-1 \choose d-1}.
\eens
Here we used the identity
\be
{d+p \choose d} - {d+p-1 \choose d} = {d+p-1 \choose d-1},
\label{ident}
\ee
which underlies our strategy for cancelling the leading contributions to
anomalies.

\subsection{ Both free bosons and fermions }

We now combine the fields from the previous two subsections, and consider
a model with the following field content:
\begin{itemize}
\item
$n_b$ bosonic fields with weight $\la_b$.
\item
$n_f$ fermionic fields with weight $\la_f$.
\end{itemize}
Each skin degree of freedom makes the following contribution to the 
central charge:
\bes
c_b &=& 2 n_b (1 - 6\la_b + 6\la_b^2), \nle
c_f &=& - 2 n_f (1 - 6 \la_f + 6\la_f^2).
\eens
Since the bosonic skin has thickness $2$ and the fermionic skin thickness
$1$, the total central charge becomes
\bes
c_\Tot &=& c_b \bigg( {d+p\choose d} - {d+p-2\choose d} \bigg)
+ c_f \bigg( {d+p\choose d} - {d+p-1\choose d} \bigg) 
\nl
&=& (c_b + c_f) {d+p\choose d} - c_f {d+p-1\choose d} - c_b {d+p-2\choose d}.
\label{cTot}
\ees
This expression vanishes if $d=0$ and is independent of $p$ when $d=1$.
In the special case that 
\be
c_b = - c_0, \qquad c_f = 2c_0, \qquad \hbox{for some $c_0 > 0$},
\label{condfree}
\ee
we find by repeated use of the identity (\ref{ident}) that the central 
charge becomes
\bes
c_\Tot &=& c_0 {d+p\choose d} - 2 c_0 {d+p-1\choose d} 
+ c_0{d+p-2\choose d} \nl
&=&	 c_0 {d+p-1\choose d-1} - c_0 {d+p-2\choose d-1} \\
&=&	 c_0 {d+p-2\choose d-2}.
\eens
In particular, when $d = 2$ the central charge $c_\Tot = c_0$ independent
of $p$; the anomaly does not diverge in the $p\to\infty$ limit.

This example exhibits the main characteristics of QJT. A priori, jet
quantization of free fields only makes sense in $d \leq 1$ dimensions,
due to the appearence of infinite reparametrization anomalies. However,
with a clever choice of field content, the leading divergencies can be
made to cancel, leaving a finite anomaly also in $d = 2$ dimensions.
In contrast, the anomaly is never convergent if $d \geq 3$; there are
simply not enough terms in (\ref{cTot}) to arrange sufficient 
cancellation.

Note that the condition $c_0 > 0$ is a necessary (but presumably not
sufficient) condition for unitarity. Equation (\ref{condfree}) leads to
a finite central charge also if $c_0 < 0$, but we also demand that the
representation of the Virasoro algebra (\ref{Vir}) is unitary, something
which is only possible if $c_\Tot > 0$.

\subsection{ Green's functions and anomalies }
\label{ssec:Green}

In this subsection we emphasize the relation between reparametrization
anomalies and locality, in the sense of Green's functions depending on
separation.

Consider some scalar field $\phi(\xx,x^0)$. The behaviour of the
correlation	function
\be
G(\xx-\yy, x^0-y^0) = \bra0 \phi(\xx,x^0)\phi(\yy,y^0)\ket0
\ee
when the physical points $x$ and $y$ coalesce is governed by 
the anomalous dimensions $h$:
\be
G(\xx-\yy, x^0-y^0) \approx {A\/((x^0-y^0)^2 - (\xx-\yy)^2)^h},
\ee
for some constant $A$. In particular, 
\be
G(\zero, x^0-y^0) \approx {A\/(x^0-y^0)^{2h}}.
\ee
The physical time coordinates are related to gauge time parameters as in
(\ref{par}): $x^0 = q^0(t)$, $y^0 = q^0(t')$. The correlation function
expressed in terms of the time parameter thus diverges when $t \to t'$
as
\be
G(\zero, t-t') \approx {B\/(t-t')^{2h}},
\label{Gtt}
\ee
where $B = A/(\dot q^0(t))^{2h}$. Since $q^0(t)$ relates physical time to
parameter time, it must be an everywhere smooth and invertible function,
which means that $\dot q^0(t) \neq 0$ for every $t$. Hence the divergence
of the Green's function is governed by the same anomalous dimension $h$,
independent of whether it is expressed in terms of physical or gauge time.

As is well known in conformal field theory, correlators of the form
(\ref{Gtt}) are compatible with local diffeomorphism symmetry on the
circle, but only if the central charge in the Virasoro algebra (\ref{Vir})
is nonzero, and indeed positive. This is because all unitary,
quantum representations of the Virasoro algebra with lowest $L_0$ 
eigenvalue $h > 0$ have $c > 0$. Hence locality and unitarity together
imply that the reparametrization symmetry be anomalous.

\section{ Gauge theory }

\subsection{ Free Maxwell field }

The action reads
\be
S = {1\/4} \int \dx\ F_\mn F^\mn,
\label{SMax}
\ee
where the field strength
\be
F_\mn = \d_\mu A_\nu - \d_\nu A_\mu.
\ee
The equations of motion,
\be
\EE_\mu = \d^\nu F_\nm = \Box A_\mu - \d_\mu \d^\nu A_\nu,
\ee
are redundant and do hence not completely fix the time evolution, due to
the identity
\be
\d^\mu\EE_\mu \equiv 0.
\ee
Consequently, the theory is invariant under the gauge symmetry
$A_\mu \mapsto A_\mu + \d_\mu X$, for $X(x)$ an arbitary function over
spacetime.

The gauge transformations generate the algebra $\map(d+1,u(1))$ of maps 
from $(d+1)$-dimensional spacetime to the abelian Lie algebra $u(1)$.
In terms of the smeared generators $\J_X = \int \dx X(x)J(x)$, the
bracket reads 
\be
[\J_X, \J_Y] = 0.
\label{mapu1}
\ee
The action on the fields is given by
\bes
[\J_X, A_\mu] &=& \d_\mu X, \nl
{[}\J_X, F_\mn] &=& 0, \\
{[}\J_X, \EE_\mu] &=& 0.
\eens
There are several ways to treat a gauge symmetry. For our purposes the
most convenient choice is a BV-BRST formalism analogous to the one 
introduced in subsection \ref{ssec:BV1}. To this end, we introduce a 
fermionic antifield $A^*_\mu$, a bosonic second-order antifield $\zeta$, 
and a fermionic ghost $c$. The BV-BRST operator $\delta$ is defined by
\bes
\delta c &=& 0, \nl
\delta A_\mu &=& \d_\mu c, 
\nlb{BVMax}
\delta A^*_\mu &=& \d^\nu F_\nm, \nl
\delta \zeta &=& \d^\mu A^*_\mu.
\eens
One readily checks that $\delta^2 = 0$ and that zeroth cohomology group
can be identified with the space of gauge-invariant functionals of 
$A_\mu(\xx,0)$ and $F_{\mu0}(\xx,0)$, i.e. gauge-invariant functionals of
the magnetic and electric fields at time $t = 0$.

\subsection{ Free Maxwell $p$-jets }

As in the scalar field case, we pass to $p$-jet space by expanding the
Maxwell field in a Taylor series:
\be
A_\mu(\xx,t) \approx \sum_{|\mm|\leq p} {1\/\mm!} 
 \,A_{\mu,\mm}(t)\,(\xx-\qq(t))^\mm.
\ee
We can immediately translate the field concepts above to their jet space
analogs. Field strength:
\be
F_{\mn,\mm} = A_{\nu,\mm+\hat\mu} - A_{\mu,\mm+\hat\nu}.
\ee
Equations of motion:
\bes
\EE_{\mu,\mm} &=& 
\sum_{\nu,\rho=0}^d \eta^{\nu\rho} F_{\nm,\mm+\hat\rho} \nle
&=& \sum_{\nu,\rho=0}^d \eta^{\nu\rho} \Big(
A_{\mu,\mm+\hat\nu+\hat\rho} - A_{\nu,\mm+\hat\mu+\hat\rho} \Big).
\eens
Redundancy:
\be
\sum_{\mu,\nu=0}^d \eta^\mn \EE_{\mu,\mm+\hat\nu} = 0.
\label{redJet}
\ee
Gauge transformations:
\be
A_{\mu,\mm} \mapsto A_{\mu,\mm} + \d_{\mm+\hat\mu} X(q),
\label{gaugeJet}
\ee
where we use the notation
\be
\d_\mm X(q) = \underbrace{\d_1 .. \d_1}_{m_1} 
\underbrace{\d_2 .. \d_2}_{m_2}\ ...\ 
\underbrace{\d_d .. \d_d}_{m_N} X(q(t)).
\ee

The equations of motion are of second order, but the redundancy 
condition (\ref{redJet}) involves derivatives of one order higher.
The thickness of the skin is thus three. The cleanest way to 
construct the $p$-jet phase space is to use the BV-BRST formalism.
We can immediately read off the action of the BRST operator $\delta$ on
$p$-jets from (\ref{BVMax}):
\bes
\delta c_\cm &=& 0, \nl
\delta A_{\mu,\mm} &=& c_\cmmu, 
\nlb{BVJet}
\delta A^*_{\mu,\mm} &=& \sum_{\nu,\rho=0}^d \eta^{\nu\rho}
F_{\nm, \mm+\hat\rho}, \nl
\delta \zeta_\cm &=& \sum_{\mu,\nu=0}^d \eta^\mn  A^*_{\mu,\mm+\hat\nu}.
\eens
In view of the second equation in (\ref{BVJet}), it might appear that we
need to define the ghost $c_\cm$ for all $|\mm|\leq p+1$.
However, the $p$-jet BRST operator only needs to implement the gauge
symmetry in the body of the $p$-jet $A_{\mu,\mm}$, which consists of 
$|\mm|\leq p-3$ because the skin has thickness three. It therefore 
suffices to define $c_\cm$ for $|\mm|\leq p-2$. The maximal order and the 
Grassmann parity of the various fields and antifields are listed in the
following table:
\be
\begin{array}{|c|c|c|}
\hline
\hbox{Field} & \hbox{Order} & \hbox{Parity} \\
\hline 
A_{\mu,\mm} & p & B \\
A^*_{\mu,\mm} & p-2 & F \\
c_\cm & p-2 & F \\
\zeta_\cm & p-3 & B \\
\hline
\end{array}
\label{tabMax}
\ee
In the parity column, $B$ stands for bosonic and $F$ for fermionic.

The algebra of gauge transformations acts on the $p$-jet $A_{\mu,\mm}$ as
\be
[\J_X, A_{\mu,\mm}] = \d_{\mm+\hat\mu} X(q).
\ee
It follows that $\J_X$ commutes with the field strength  $F_{\mn,\mm}$
and with the equations of motion. Therefore, the action on the fields and
anti-fields in (\ref{BVJet}) is given by 
\be
[\J_X, A^*_{\mu,\mm}] = [\J_X, c_\cm] = [\J_X, \zeta_\cm] = 0.
\ee
An explicit off-shell realization of the gauge generators, acting in the 
space of arbitary $p$-jet histories, is
\be
\J_X = \sum_{|\mm|\leq p} \sum_{\mu=0}^d 
 \int dt\ i\d_{\mm+\hat\mu} X(q(t)) E^{\mu,\mm}(t),
\label{JXMax}
\ee
where $E^{\mu,\mm} = -i\delta/\delta A_{\mu,\mm}$ satisfies the
instantaneous CCR
\be
[A_{\mu,\mm}(t), E^{\nu,\nn}(t')] = 
 i \delta^\nu_\mu \delta^\nn_\mm \delta(t-¨t').
\ee
Because the expression (\ref{JXMax}) is linear in $E$, normal ordering
is not possible and the gauge algebra does not acquire any extension.

\subsection{ Yang-Mills theory }

The situation becomes more interesting if we consider a non-abelian gauge
theory. Let $\oj$ denote a finite-dimensional Lie algebra with generators
$J^a$, totally anti-symmetric structure constants $f^{abc}$, and
Killing metric $\delta^{ab}$. The Lie brackets are given by 
\be
[J^a, J^b] = if^{abc} J^c.
\ee
The gauge transformations generate the algebra $\map(d+1,\oj)$ of maps
from $(d+1)$-dimensional spacetime into $\oj$. For every $\oj$-valued 
function $X = X_a(x)J^a$, we define the smeared operator $\J_X$, with
brackets
\be
[\J_X, \J_Y] = \J_{[X,Y]},
\label{mapg}
\ee
where $[X,Y] = if^{abc} X_a Y_b J^c$. This algebra admits the ``central''
extension \cite{PS86}
\be
[\J_X, \J_Y] = \J_{[X,Y]} 
 - {k\/2\pi i} \delta^{ab} \int \dot q^\mu(t) X_a(q(t)) \d_\mu Y_b(q(t)),
\label{KMd}
\ee
which is the natural generalization of the affine Kac-Moody algebra
$\hat\oj$ to multi-dimensions. Since this extension is proportional to the
second Casimir $\tr J^aJ^b$ rather than to the third Casimir
$d^{abc} = \tr (J^aJ^b + J^bJ^a)J^c$, it does not arise in QFT.

The construction of the phase space proceed in analogy with the abelian
case. The covariant derivative:
\be
D_\mu = \d_\mu + A^a J^a.
\ee
Field strength:
\be
F^a_\mn = [D_\mu, D_\nu]^a.
\ee
Equations of motion:
\be
\EE^a_\mu = (D^\nu F_\nm)^a = 0.
\ee
Redundancy:
\be
(D^\mu \EE_\mu)^a \equiv 0.
\ee
The construction of the  phase space is simplest within the BV-BRST 
formalism. To this end, we introduce
an antifield $A^a_\mu$, a ghost $c^a$, and a second-order antifield
$\zeta^a$. The BV-BRST operator $\delta$ acts in the extended phase space
as
\bes
\delta c^a &=& if^{abc} c^b c^c, \nl
\delta A^a_\mu &=& if^{abc} A^b_\mu c^c + \d_\mu c^a, 
\nlb{BVYM}
\delta A^{*a}_\mu &=& (D^\nu F_\nm)^a + if^{abc} A^{*b}_\mu c^c, \nl
\delta \zeta^a &=& (D^\mu A^*_\mu)^a + if^{abc} \zeta^b c^c.
\eens
The zeroth cohomology group can be identified with the space of 
gauge-invariant functionals over $A^a_\mu(\xx,0)$ and 
$\d_0 A^a_\mu(\xx,0)$, i.e. functionals over the physical phase space.

We now proceed to $p$-jets. The algebra of gauge transformations
(\ref{mapg}) acts on the fields and antifields:
\bes
[\J_X, A^a_{\mu,\mm}] &=& if^{abc} (X^b A^c_\mu)_\cm 
+ \d_{\mm+\hat\mu} X^a, \nl
{[}\J_X, A^{*a}_{\mu,\mm}] &=& if^{abc} (X^b A^{*c}_\mu)_\cm, 
\nlb{JXYM}
{[}\J_X, c^a_\cm] &=& if^{abc} (X^b c^c)_\cm, \nl
{[}\J_X, \zeta^a_\cm] &=& if^{abc} (X^b \zeta^c)_\cm,
\eens
where $(X A)_\cm$ denotes the $\mm$:th coefficient in the Taylor expansion
of \break $X(\xx) A(\xx,t)$ around the point $\xx = \qq(t)$, i.e.
\be
(X A)_\cm = \sum_{|\nn|\leq p} {\mm\choose\nn} 
 \d_{\mm-\nn} X(\qq(t)) A_\cn(t).
\ee
From (\ref{JXYM}) we can read off an explicit expression for the
gauge generators $\J_X$, in analogy with (\ref{JXMax}). However, in
constrast to the abelian case, this expression contains bilinear terms
which must be normal ordered after quantization. This normal ordering
gives rise to a Kac-Moody-like extension of the form (\ref{KMd}).

If a field transforms in the $\oj$ representation $\varrho$, the 
``abelian charge'' $k$ in (\ref{KMd}) is $k = - Q_\varrho$ (if the field
is bosonic) or $k = + Q_\varrho$ (if the field is fermionic), where 
the value of the second Casimir operator in $\varrho$ is given by
\be
\tr J^a J^b = Q_\varrho \delta^{ab}.
\label{Casimir2}
\ee
In particular, the gauge potential and its antifields and ghost all 
transform in the adjoint representation of $\oj$, so $\varrho = \ad$
here. Moreover, $A^a_\mu$ and $A^{*a}_\mu$ are vector fields with 
$d+1$ components. We can therefore write down a list analogous to
(\ref{tabMax}), with an extra column which denotes the contribution to
the abelian charge from each jet component.
\be
\begin{array}{|c|c|c|c|}
\hline 
\hbox{Field} & \hbox{Order} & \hbox{Parity} & k \\
\hline 
A^a_{\mu,\mm} & p & B & -(d+1) Q_\ad \\
A^*_{\mu,\mm} & p-2 & F & (d+1) Q_\ad \\
c_\cm & p-2 & F & Q_\ad \\
\zeta_\cm & p-3 & B & -Q_\ad \\
\hline
\end{array}
\label{tabYM}
\ee
The total extension is thus
\bes
k_\Tot &=& -(d+1) Q_\ad {d+p \choose d} +(d+1) Q_\ad {d+p-1 \choose d} \nl
&&\quad +\ Q_\ad {d+p-2 \choose d} - Q_\ad {d+p-3 \choose d} \\
&=& -(d+1) Q_\ad {d+p-1 \choose d-1} + Q_\ad {d+p-3 \choose d-1}.
\eens
In the non-abelian case, $k_\Tot$ vanishes if $d = 0$, it equals $-Q_\ad$
independent of $p$ if $d = 1$, and it diverges with $p$ if $d \geq 2$. In
the abelian case, $Q_\ad = 0$, and $k_\Tot = 0$ in any numbers of 
dimensions, in agreement with our observation in the previous subsection.

\subsection{ Matter fields }

We introduce fermions via the minimal coupling prescription. For
simplicity, we write down formulas for the Maxwell theory only, but it is
straightforward to write down the required modifications in the non-abelian 
case.  To the free Maxwell action (\ref{SMax}) we add the Dirac action 
\be
S_\psi = \int \dx\ 
\bar\psi\Big( i\gamma^\mu(\d_\mu - eA_\mu) - \ww\Big)\psi.
\ee
The equations of motion become
\bes
\EE^\mu_A &=& \d_\nu F^\nm - j^\mu = 0, \nl
\EE_\psi &=& i\gamma^\mu(\d_\mu - eA_\mu)\psi - \ww\psi = 0, \\
\bar\EE_\psi &=& i(\d_\mu + eA_\mu)\bar\psi\gamma^\mu + \ww\bar\psi = 0,
\eens
where the current
\be
j^\mu = i e\bar\psi\gamma^\mu\psi.
\ee
Because of current conservation, $\d_\mu j^\mu = 0$, the equations of
motion are redundant and do not completely fix the time evolution:
\be
\d^\mu\EE_\mu \equiv 0.
\ee
As a result, we still have an $u(1)$ gauge symmetry (\ref{mapu1}), which
acts on the fields as
by
\bes
[\J_X, A_\mu] &=& \d_\mu X, \nl
{[}\J_X, \psi] &=& eX\psi, \\
{[}\J_X, \bar\psi] &=& -eX\psi.
\eens
In $p$-jet space, the action on the fermions reads 
$[\J_X, \psi_\cm] = (X\psi)_\cm$, etc.

Bosonic matter can also be introduced, e.g. by adding a scalar 
electrodynamics term (\ref{SAphi}) to the action. For our purposes, the
main difference between bosons and fermions resides in the sign of the
extension (\ref{KMd}). If we assume that there are $n_\phi$ bosonic
species $\phi$ and $n_\psi$ fermionic species $\psi$ (where the 
conjugate $\bar\psi$ counts as another species), the bosonic and
fermionic contributions to the abelian charge $k$ become
\be
k_\phi = - n_\phi Q_\phi, \qquad
k_\psi = + n_\psi Q_\psi,
\ee
respectively. We can readily generalize this to several different type
of species transforming in different representations $\varrho$; the
abelian charge is simply the sum of the contributions from the different
species, including sign.

We now turn to a general non-abelian gauge theory, with both fermionic and 
bosonic matter. After passage to $p$-jet space, the BV-BRST complex is
built from the following content:
\be
\begin{array}{|c|c|c|c|}
\hline 
\hbox{Field} & \hbox{Order} & \hbox{Parity} & k \\
\hline 
A^a_{\mu,\mm} & p & B & -(d+1) Q_\ad \\
A^*_{\mu,\mm} & p-2 & F & (d+1) Q_\ad \\
c_\cm & p-2 & F & Q_\ad \\
\zeta_\cm & p-3 & B & -Q_\ad \\
\psi_\cm & p & F & n_\psi Q_\psi \\
\psi^*_\cm & p-1 & B & -n_\psi Q_\psi \\
\phi_\cm & p & B & -n_\phi Q_\phi \\
\phi^*_\cm & p-2 & F & n_\phi Q_\phi \\
\hline
\end{array}
\label{tabYMD}
\ee
The columns list the type of field, the maximal order for which the 
corresponding jet is  defined, its Grassmann parity (bose/fermi) and the 
contribution to the abelian charge in (\ref{KMd}). Summing the various 
contributions, the total abelian charge is
\[
k_\Tot = k_0 {d+p \choose d} + k_1 {d+p-1 \choose d} 
+ k_2 {d+p-2 \choose d} + k_3 {d+p-3 \choose d},
\]
where 
\bes
k_0 &=& -(d+1) Q_\ad + n_\psi Q_\psi - n_\phi Q_\phi, \nl
k_1 &=& - n_\psi Q_\psi, 
\nlb{k0k3}
k_2 &=& (d+1) Q_\ad + Q_\ad + n_\phi Q_\phi, \nl
k_3 &=& - Q_\ad.
\eens
If we choose 
\be
k_1 = -3 k_0, \qquad k_2 = 3 k_0, \qquad k_3 = -k_0,
\label{kkk}
\ee
repeated use of the identity (\ref{ident}) leads to
\be
k_\Tot = k_0 {d+p-3 \choose d-3}.
\ee
Provided that the conditions (\ref{kkk}) are satisfied, the total abelian 
charge vanishes when $d \leq 2$ and has a finite value if $d = 3$. 
We read off from (\ref{k0k3}) that in order for this to happen, 
we must have
\bes
-(d+1) Q_\ad + n_\psi Q_\psi - n_\phi Q_\phi &=& k_0, \nl
- n_\psi Q_\psi &=& -3k_0, \nle
(d+1) Q_\ad + Q_\ad + n_\phi Q_\phi &=& 3k_0, \nl
- Q_\ad &=& -k_0.
\eens
These are four equations for three unknowns, and would in general not
be solvable. However, the equation system turns out to be singular, and
admits the solution
\bes
k_0 &=& Q_\ad, \nl
n_\psi Q_\psi &=& 3Q_\ad, 
\label{YMcond}\\
n_\phi Q_\phi &=& (1-d)Q_\ad\ =\ - 2Q_\ad, \qquad \hbox{if $d = 3$.}
\eens
In order for the QJT of a non-abelian gauge theory to have a finite
gange anomaly in $3+1$ dimensions, these conditions on the matter content 
are necessary. 

No interesting solution to the conditions (\ref{YMcond}) has been found, 
and in fact there is a serious problem with the negative sign in the last
equation; it implies a violation of the spin-statistics problem. Since
$\phi$ is bosonic, the abelian charge $k_\phi = - n_\phi Q_\phi$ must be
negative; however, the last equation above implies that $k_\phi$ is 
positive, so $\phi$ must in fact be fermionic. On the other hand, we 
assumed that the equations of motion for $\phi$ are second order, which
implies that $\phi$ has integer spin. The only solution to (\ref{YMcond})
is thus that $\phi$ is an integer spin fermion, which violates the 
spin-statistics theorem if $\phi$ is a physical field. We discuss this
matter further in the conclusion.

\section{ Conclusion }

In this paper we considered QJT as a regularization method: we 
replace all fields by $p$-jets, i.e. their Taylor expansions truncated at
order $p$. Although the space of $p$-jets is finite-dimensional, the 
$p$-jet phase space is infinite-dimensional, because only histories in the
body are specified by initial conditions. 

The $p$-jet regularization does not preserve the gauge symmetries of the
original theory. The gauge symmetries become anomalous in the regularized
theory due to the infinite dimensionality of the skin, and this anomaly
does not vanish when the regularization is removed by taking the $p \to
\infty$ limit. Worse, the corresponding ``abelian charges'' diverge in
more than $1+1$ dimensions, but with a clever choice of field content the
anomalies can be rendered finite in $3+1$ dimension; the critical number
of spatial dimensions $d = 3$ equals the thickness of the skin.

We studied conditions for cancelling the divergent parts of Yang-Mills
anomalous, but no solutions were found. In fact, the solution in
(\ref{YMcond}) appears to violate the spin-statistics theorem; the field
$\phi$ should be fermionic but have second-order equations of motion, i.e.
integer spin. It is conceivable that one could interpret $\phi$ as the
ghost for some additional symmetry, perhaps having something to do with
confinement or gauge symmetry breaking. If so, the apparent violation of
the spin-statistics theorem is not a problem, because $\phi$ is not a
physical field. This issue deserves further investigation.

\end{document}